# Statistics of Coulomb Blockade Peak Spacings

S. R. Patel, S. M. Cronenwett, D. R. Stewart, A. G. Huibers, and C. M. Marcus
*Department of Physics, Stanford University, Stanford, CA 94305*

C. I. Duruöz and J. S. Harris Jr.
*Electrical Engineering Department, Stanford University, Stanford, CA 94305*

K. Campman and A. C. Gossard
*Materials Department, University of California at Santa Barbara*
*Santa Barbara, CA 93106*

Distributions of Coulomb blockade peak spacing are reported for large ensembles of both unbroken (magnetic field $B = 0$) and broken ($B \neq 0$) time reversal symmetry in GaAs quantum dots. Both distributions are symmetric and roughly gaussian with a width ~ 2–6% of the average spacing, with broad, non-gaussian tails. The distribution is systematically wider at $B = 0$ by a factor of ~ $1.2 \pm 0.1$. No even-odd spacing correlations or bimodal structure in the spacing distribution is found, suggesting an absence of spin-degeneracy. There is no observed correlation between peak spacing and peak height.

For some time it has been appreciated that electron transport in mesoscopic systems exhibits quantum interference effects with universal statistical features, and that this universality can be associated with the underlying universality of quantum chaos [1] and its mathematical description in terms of random matrix theory (RMT) [2]. This approach has been successful in describing transport in open quantum systems (i.e. systems with large conductance, $g > e^2/h$, to reservoirs) where a single-particle picture apparently provides an adequate description of the physics. Recent application of RMT to ground state properties of nearly isolated quantum dots—in particular, in characterizing the distributions of Coulomb blockade (CB) conductance peak heights [3] has also been remarkably successful [4].

On the other hand, experiments by Sivan *et al.* [5] and Simmel *et al.* [6] suggest that the most basic prediction of RMT, namely the famous Wigner surmise for the distribution of level spacings, fails to describe the fluctuations of Coulomb blockade peak spacing, implying that fluctuations in the energy separation between adjacent ground states of a quantum dot—the so-called addition spectrum—appear not to be distributed according to RMT. In particular, these experiments [5, 6] found CB peak spacing fluctuations of order 0.1–0.15 of the average spacing, larger than predicted by RMT assuming constant charging energy $E_C = e^2/C_{dot}$ with the total capacitance of the dot given by $C_{dot}$. The large fluctuations observed in the experiment and supporting numerics lead Sivan *et al.* [5] to suggest that *classical* charging energy fluctuations proportional to $E_C$ not included in RMT dominate peak spacing fluctuations. On the other hand, recent random phase approximation (RPA) calculations lead to the opposite conclusion, that fluctuations due to charge rearrangement should be smaller than [7] or of order [8] the mean single-particle level spacing $\Delta$, much smaller than $E_C$. However, one should use caution in applying these theories to semiconductor quantum dots, since RPA breaks down at low electron densities where single-particle kinetic energies and interparticle potential energies are comparable.

In this Letter, we present an extensive experimental study of the spacings of CB peaks in GaAs quantum dots, for both zero and non-zero magnetic field, including over 20,000 CB peaks measured in seven devices. We find that the distributions of CB peak spacing fluctuations are not well described by single-particle, *spin-degenerate* RMT (SDRMT); the main discrepancy being the absence of bimodal structure in the measured distributions. Both B = 0 and B $\neq$ 0 peak spacing distributions are roughly gaussian with non-gaussian tails, in qualitative agreement with [5, 6]. In contrast to previous experiments, however, we find that the width of the peak spacing distribution is narrow, comparable (once scaled) to the mean level spacing, and shows the effects of time reversal symmetry breaking comparable to RMT predictions, suggesting that quantum effects play a role in determining the distributions. We also find no correlation between CB peak heights (reflecting eigenfunction properties) and peak spacings (reflecting eigenvalue properties).

Ground state energy fluctuations are measured using the Coulomb blockade, which appears in quantum dots with tunneling leads (left, right lead conductance $g_l, g_r < 2e^2/h$), when temperature T and source-drain voltage $V_{sd}$ are less than $E_C$ [9, 10]. In this regime, dot conductance is suppressed except when the ($N$+1) and $N$ electron ground state energies of the dot differ by the chemical potential of the leads; this degeneracy condition, when tuned by a gate voltage $V_g$, produces a series of narrow conductance peaks nearly periodic in $V_g$. At very low temperature and bias, ($k_B$T, $eV_{sd}$) ~< $\Delta$ (we define $\Delta = 2\pi\hbar^2/m^*A$ as the average single particle level spacing for spin-degenerate levels; $m^*$ is the electron effective mass, and $A$ is the area occupied by the electrons), transport on a CB peak is a resonant process, making the peak position sensitive to the discrete spectrum of the dot. A simple model connecting level spacing statistics and CB peak spacing is the so-called 'constant interaction' (CI) model, in which the separation between ground state energies of the dot probed by CB is separated into two parts, a charging energy $E_C$ that is independent of (or at most slowly varying with) the number of electrons on the dot, *N*, and a fluctuating part associated with a discrete quantum level spacing. This separation assumes that fluctuations in $E_C$ due to charge rearrangement in the ground state upon adding an electron are small compared to Δ. Within the CI model, the spacing (in gate voltage) between CB peaks, $\Delta V_g^i = V_g^{i+1} - V_g^i$, where $V_g^i$ is the center position of the $i^{\text{th}}$ peak, is given by

$$e\eta\Delta V_g^i = \begin{cases} (\varepsilon^{i+1} - \varepsilon^i) + E_C & (i \text{ even}) \\ E_C & (i \text{ odd}) \end{cases} \tag{1}$$

where $\varepsilon^i$ is the $i^{\text{th}}$ single-particle energy level and $\eta = C_g / C_{dot}$ is the ratio of gate capacitance to total dot capacitance, also assumed to vary slowly with *i*. The dependence of Eq. 1 on whether *i* is even or odd reflects the spin degeneracy of levels.

If one further assumes that spacings $(\varepsilon^{i+1} - \varepsilon^i)$ obey spin-degenerate (SD) RMT statistics, the resulting distribution $P(\nu)$ of normalized fluctuations in peak spacing $\nu$ = $\left(\Delta V_g^i - \langle \Delta V_g^i \rangle\right) / \langle \Delta V_g^i \rangle$ will consist of a $\delta$ function for the case of *i* odd plus a Wigner-Dyson distribution ($P_{WD}(s) \propto s\, e^{-\pi s^2/4}$ for B = 0 and $\propto s^2\, e^{-4s^2/\pi}$ for B ≠ 0, with $s$ in units of $\Delta$) for the case *i* even. The brackets $\langle \cdot \rangle$ denote an average over an ensemble of peaks. Note that within this CI+SDRMT model, the standard deviation, $\sigma$, of $P(\nu)$ depends on time-reversal symmetry; for $\Delta << E_C$, CI+SDRMT gives $\sigma = 0.62(0.58)\Delta / E_C$ for B = 0 (B ≠ 0) and a ratio $\sigma_{B=0} / \sigma_{B\neq 0} \sim 1.1$. For GaAs dots of the type described here and in Ref. [5, 6], the predicted peak spacing distribution width is $\sigma(\nu) \sim 0.03$, depending slightly on the dot shape and size, based on experimental values of $\Delta$ and $E_C$ (see Table 1).

Since a bimodal $P(\nu)$ is not found experimentally, as seen below, one is motivated to consider a CI model without even-odd structure, based on *spin-resolved* (SR) level statistics. The breaking of spin degeneracy, discussed in [7] and observed experimentally in [11], will lead to level statistics described roughly by two overlapping Wigner-Dyson distributions [12]. The ratio $\sigma_{B=0}/\sigma_{B\neq 0} = 0.70/0.65 \sim 1.1$ in CI+SRRMT is similar to CI+SDRMT; however, the predicted spacing distribution width $\sigma(\nu) \sim 0.01$ is reduced by a factor of ~ 2 due to a reduced mean level spacing, $\Delta_{SR} = \Delta/2$.

The quantum dots were formed by gate depletion of a two-dimensional electron gas (2DEG) in a GaAs/AlGaAs heterostructure (see Table 1). Irregular dot shapes were designed to produce chaotic scattering, and all devices were smaller than the bulk mean free path so that transport within the dots is ballistic. The dot area, $A$, was estimated from the lithographic area with a depletion of ~120 nm around the perimeter of the device. Charging energy was measured from the relation $E_C \sim e\eta\langle\Delta V_g\rangle$, based on Eq. 1 with $\Delta << E_C$, with $\eta$ extracted from the T dependence of the peak width [10]. All measurements were made using two-wire ac lockin techniques with a voltage bias of 5 μV at 11 Hz. The electron base temperature, determined by fitting peak width versus temperature [10], was ~ 100 mK for all devices. Ensemble statistics were collected by sweeping one gate voltage, $V_g$, over ~ 20 peaks then incrementing magnetic field or a second gate voltage to yield a new ensemble of peaks. All data are based on relatively small CB peaks, in the range 0.01 to 0.1 $e^2/h$.

A typical scan of CB peaks is shown in Fig. 1a. To extract peak spacing, each peak is fit by a $cosh^{-2}$ form [10] and spacing is determined from the centers of the fits (Fig. 1b). Since the average spacing decreases with increasing $N$ (reflecting an increasing $C_{dot}$), a running average $\langle\Delta V_g^i\rangle$ is found from the best fit line (dashed in Fig. 1b) and used to define $\nu = \left(\Delta V_g^i - \langle\Delta V_g^i\rangle\right)/\langle\Delta V_g^i\rangle$. Experimental noise in the spacing distribution (for instance due to charge rearrangement in the doping layer), given as $\sigma_{Noise}(\nu)$ in Table 1, can be separated from real spacing fluctuations by comparing measurements at opposite magnetic fields (Fig. 1c). The noise $\sigma_{Noise}(\nu)$ is defined as the distribution width of fluctuations antisymmetric in B and so would not include field independent gate voltage induced charge rearrangement. While the heights of nearby CB peaks show considerable correlation (Fig. 1a), peak spacings appear uncorrelated (Fig. 1b, 1c). We estimate the number of "independent" peak spacings, $n_i$ in Table 1, as the number of peaks measured in each scan of $V_g$ multiplied by the number of peak scans with characteristically different heights.

Histograms of peak spacings from three dots with similar device parameters (dots 3, 4, and 5) are shown in Fig. 2. Both the B = 0 and B ≠ 0 histograms are roughly symmetric and gaussian. A gaussian fit to the B = 0 histogram gives a standard deviation $\sigma_{B=0}(\nu)_f = 0.019$ (the subscript *f* indicates 'gaussian fit') whereas a direct evaluation of the second moment of the spacing data set yields $\sigma_{B=0}(\nu) = 0.027$. This difference results from broad non-gaussian tails, which can be seen a logarithmic plot (right insets, Fig. 2). All B ≠ 0 data are taken with between $3-15\varphi_0$ through the device, where $\varphi_0 = h/e = 4.14$ mT μm². The B ≠ 0 distribution width from a gaussian fit yields $\sigma_{B\neq0}(\nu)_f = 0.015$ and the second moment of the data gives $\sigma_{B\neq0}(\nu) = 0.022$. By either method, a ratio $\sigma_{B=0}/\sigma_{B\neq0}$ ~ 1.2–1.3 is obtained. From all devices, $\sigma_{B=0}/\sigma_{B\neq0} = 1.2 \pm 0.1$ (see Table 1), slightly larger than but comparable to the value predicted by either CI+SDRMT or CI+SRRMT. The left insets in Fig. 2 show that the CI+SDRMT spacing distributions for these devices convolved with (gaussian) experimental noise remain bimodal, clearly inconsistent with the data.

The B ≠ 0 peak spacing histogram for the smallest, quietest device (#1) is shown in Fig. 3a along with a no-adjustable-parameters CI+SRRMT distribution convolved with experimental noise. An important consideration is how the width of this distribution is affected by temperature. Figure 3b shows $\sigma_{B\neq 0}\left(k_B \mathrm{T}/\Delta_{SR}\right)$ measured for two gate voltage configurations of device 1 (with different $\Delta_{SR}$'s) along with a simple model based on thermal averaging of the peak spacings. In the model, finite-temperature peak spacings $y_i = \sum_j w_{i,j}\, s_j$ are constructed from zero-temperature spacings $s_i = (\varepsilon^{i+1} - \varepsilon^i)$ obeying SRRMT statistics with $\langle s_i \rangle = \Delta_{SR}$. The weighting function $w_{i,j} \propto cosh^{-2}(|\varepsilon^i - \varepsilon^j|/2k_B \mathrm{T})$ is chosen to include more contributions from adjacent spacings $s_i$ at higher temperature (not including height fluctuations) leading to $\sigma(\mathrm{T}) \propto \mathrm{T}^{-1/2}$ in the limit $k_B\mathrm{T} >> \Delta_{SR}$, consistent with the data. While this model appears to have the correct cross-over scale and correct limits at $k_B\mathrm{T} << \Delta_{SR}$ and $k_B\mathrm{T} >> \Delta_{SR}$, a proper theory of $\sigma(\mathrm{T})$ has yet to be developed. However, the observed temperature dependence suggests a quantum rather than classical origin to the spacing fluctuations.

Finally, we have looked for correlations between peak spacing fluctuations and peak height fluctuations. Within RMT, no correlations are expected since height fluctuations depend on eigenfunctions while spacing fluctuations depend on eigenvalues which are uncorrelated with eigenfunctions [13]. Experimentally, we find that the normalized spacing $\nu$ between adjacent peaks ($i$+1 and $i$) is not correlated with either the normalized height difference, $\Delta\tilde{g} = (g^{i+1}_{\max} - g^{i}_{\max})/\langle g_{\max}\rangle$, or their normalized average height, $\tilde{g} = \left(g^{i+1}_{\max} + g^{i}_{\max}\right)/2\langle g_{\max}\rangle$, as shown in Fig. 3c and 3d.

We thank D. Sprinzak for numerous contributions to the experiment. Also, we thank O. Agam, I. Aleiner, Y. Alhassid, B. Altshuler, A. Andreev, A. Kamenev, A. Mirlin, U. Sivan, and M. Stopa for valuable discussions. Work at Stanford supported by the ARO, ONR-YIP, the NSF-NYI, PECASE, and A. P. Sloan Foundation (Physics); JSEP (Electrical Engineering); the NSF Graduate Fellowship (S.M.C.); and the Hertz Foundation (A.G.H.). Work at UCSB supported by the AFOSR and QUEST.

## Captions

Fig. 1. (a) Coulomb blockade peaks (diamonds) at B = 30 mT as a function of gate voltage $V_g$ for device 1. Solid curve shows fits to *cosh*$^{-2}$ lineshape. Left inset: Detailed view of data and fit on log-linear scale. Right inset: Micrograph of device 1, other devices are similar. (b) Peak spacings extracted from data in (a) at B = +30 mT (diamonds) B = -30 mT (open circles). Dashed line is best fit (to +30 mT data), corresponding to $\left\langle \Delta V_g^i \right\rangle$. (c) Dimensionless peak spacing fluctuations, $\nu = \left(\Delta V_g^i - \left\langle \Delta V_g^i \right\rangle\right) / \left\langle \Delta V_g^i \right\rangle$, as a function of gate voltage $V_g$ for data in (b). Differences between ±30 mT data indicated experimental noise. Normalized (spin-resolved) mean level spacing $\Delta_{SR} / E_C$ indicated by vertical bar (See Table 1).

Fig. 2. Histograms of normalized peak spacing ν (bars) for (a) B = 0 and (b) B ≠ 0 for devices 3, 4, and 5. Solid curves show best fit to normalized gaussian of width 0.019 (0.015) for B = 0 (B ≠ 0). The B = 0 histogram is wider by a factor of ~ 1.2 than the B ≠ 0 histogram. Data represents 4,300 (10,800) CB peaks from the devices with ~ 720 (1600) statistically independent for B = 0 (B ≠ 0). Horizontal bar indicates (spin-resolved) mean level spacing $\Delta_{SR} / E_C$ averaged over the 3 devices. Right insets: Plots of histogram (diamonds) and best fit gaussian (solid curve) on log-linear scale. Dashed curve is gaussian of width 0.13 from Ref. [5]. Left insets: Dotted curves are CI+SDRMT peak spacing distributions; solid curves correspond to CI+SDRMT distributions convolved with gaussian of width $\sigma_{Noise}(\nu) = 0.009$ averaged over the three dots (see Table 1).

Fig. 3. (a) Histogram of peak spacing data from dot 1 at B ≠ 0 (bars). Dotted curve shows zero-temperature B ≠ 0 CI+SRRMT peak spacing distribution; solid curve is CI+SRRMT distribution convolved with gaussian of width $\sigma_{Noise}(\nu) = 0.006$. (b) Temperature dependence of $\sigma_{B\neq 0}(\nu)$ for two gate configurations of dot 1 ($\Delta_{SR} = 21\,\mu\text{eV}$ solid diamonds; $\Delta_{SR} = 14\,\mu\text{eV}$ open circles), and thermal averaging model (solid curve) described in text, with $\Delta_{SR} = 21\,\mu\text{eV}$ and $E_C = 590\,\mu\text{eV}$. (c),(d) Scatter plot of normalized peak spacing $\nu$ versus adjacent peak height differences $\Delta\tilde{g} = (g_{\max}^{i+1} - g_{\max}^{i})/\langle g_{\max}\rangle$ and averages $\tilde{g} = \left(g_{\max}^{i+1} + g_{\max}^{i}\right)\big/2\langle g_{\max}\rangle$ for all B ≠ 0 data from dot 4. No correlations between spacings and average heights or height differences are observed.

| Parameter | dot 1 | dot 2 | dot 3 | dot 4 | dot 5 | dot 6 | dot 7 |
|---|---|---|---|---|---|---|---|
| $A$ (μm$^2$) | 0.17 | 0.20 | 0.32 | 0.34 | 0.38 | 0.47 | 0.50 |
| d (Å) | 900 | 900 | 800 | 800 | 900 | 800 | 900 |
| $\Delta_{SR}$ (μeV) | 21 | 18 | 11 | 11 | 9 | 8 | 7 |
| $E_C$ (μeV) | 590 | 760 | 580 | 500 | 380 | 600 | 320 |
| $N$ | 340 | 400 | 900 | 1000 | 800 | 1400 | 1000 |
| $n_i$ | 190 | 70 | 140 | 830 | 1300 | 710 | 420 |
| $\sigma_{B=0}(\nu)$ ($\times 10^{-3}$) | -- | 48 (7) | 38 (2) | 25 (2) | 25 (3) | 43 (2) | 56 (3) |
| $\sigma_{B\neq 0}(\nu)$ ($\times 10^{-3}$) | 18 (2) | 34 (4) | 23 (3) | 22 (1) | 20 (2) | 38 (2) | 43 (2) |
| $\frac{\sigma_{B=0}(\nu)}{\sigma_{B\neq 0}(\nu)}$ | -- | 1.3 (0.2) | 1.7 (0.2) | 1.2 (0.1) | 1.2 (0.2) | 1.1 (0.1) | 1.3 (0.1) |
| $\sigma_{Noise}(\nu)$ ($\times 10^{-3}$) | 6 | 23 | 15 | 10 | 8 | 25 | 30 |

Table 1. Device parameters and measured spacing statistics at $T$ ~ 100 mK: area occupied by electrons ($A$), 2DEG depth (d), SR mean level spacing ($\Delta_{SR} = \pi\hbar^2 / m^* A$), charging energy ($E_C = e^2 / C_{dot}$), number of electrons in dot ($N$), number of statistically independent peak spacings ($n_i$), and peak spacing distribution width at B = 0 ($\sigma_{B=0}(\nu)$) [B ≠ 0 ($\sigma_{B\neq 0}(\nu)$)] with uncertainties in parentheses. An estimate of the noise in each data set is given as the width of the noise distribution, $\sigma_{Noise}(\nu)$; devices 2, 6 and 7 exhibit a correlation between enhanced $\sigma_{Noise}(\nu)$ and larger $\sigma(\nu)$. Devices 1, 2, 5, and 7 (3, 4, and 6) have a sheet density $n_s \sim 2\times 10^{11}\ cm^{-2}$ ($3\times 10^{11}\ cm^{-2}$) and mobility $\mu \sim 1.4\times 10^5\ cm^2/Vs$ ($6.5\times 10^5\ cm^2/Vs$).

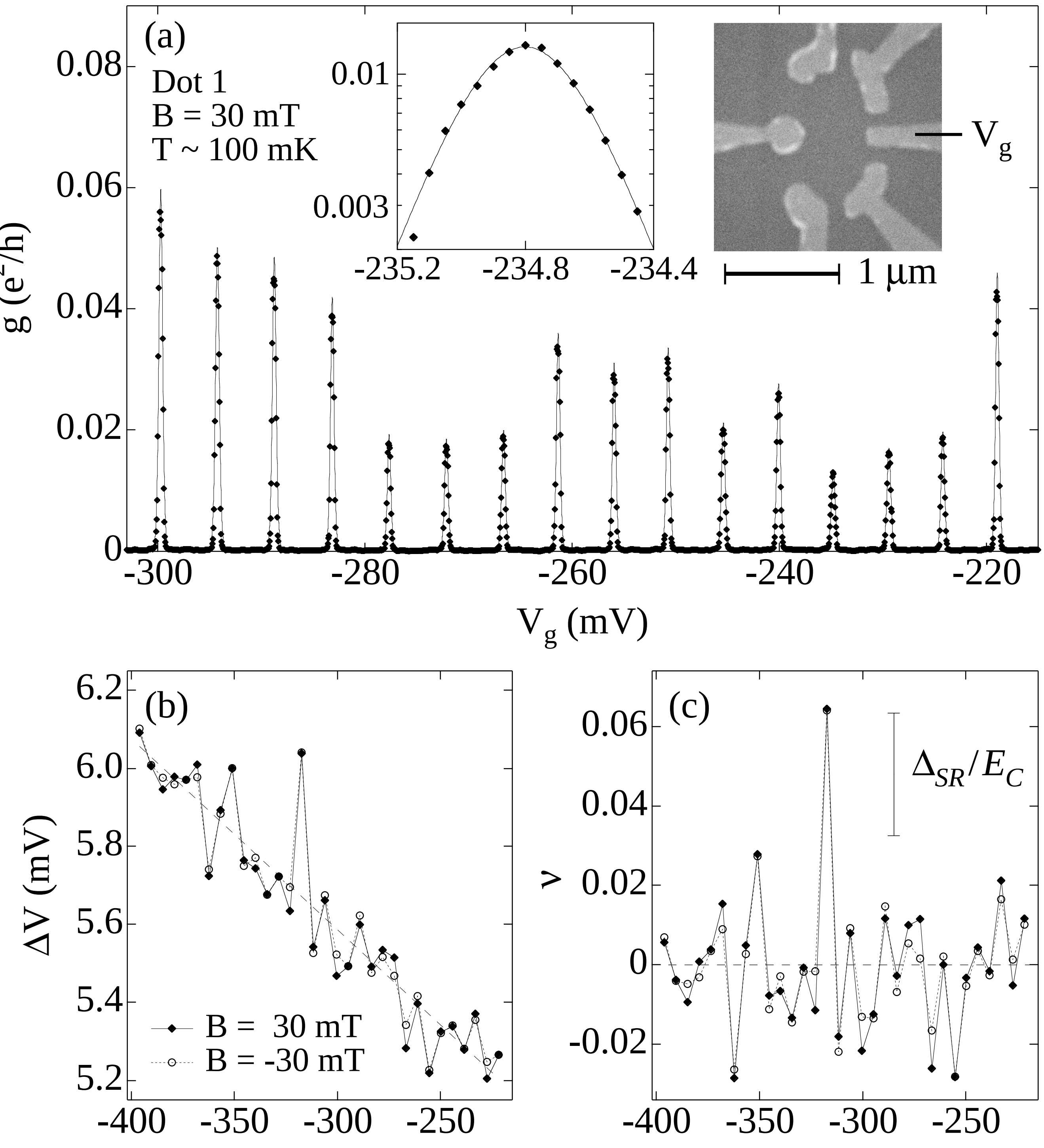

Fig. 1

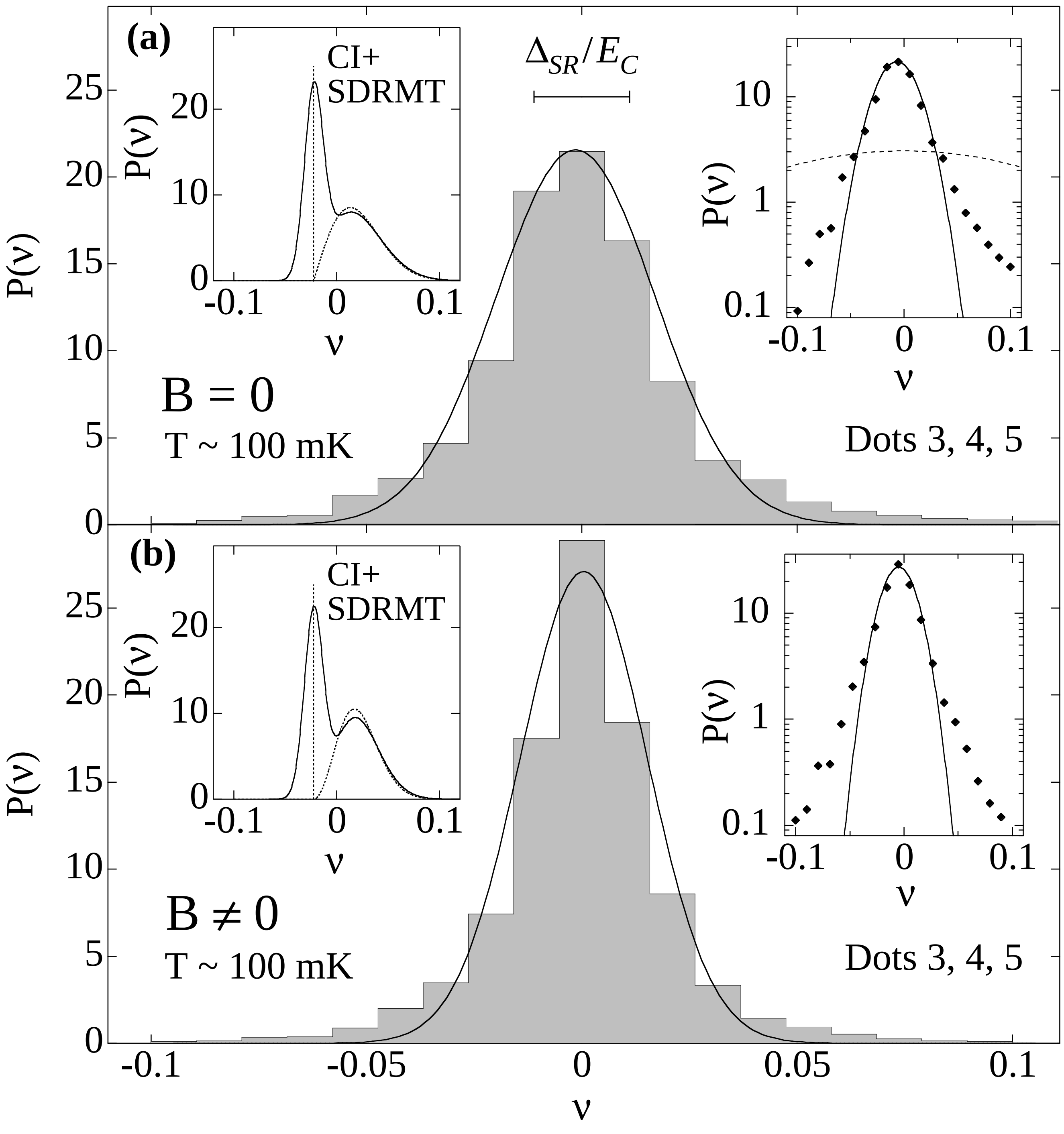

(a)
CI+
SDRMT
$\Delta_{SR}/E_C$
B = 0
T ~ 100 mK
Dots 3, 4, 5
(b)
CI+
SDRMT
B ≠ 0
T ~ 100 mK
Dots 3, 4, 5
P(ν)
ν

Fig. 2

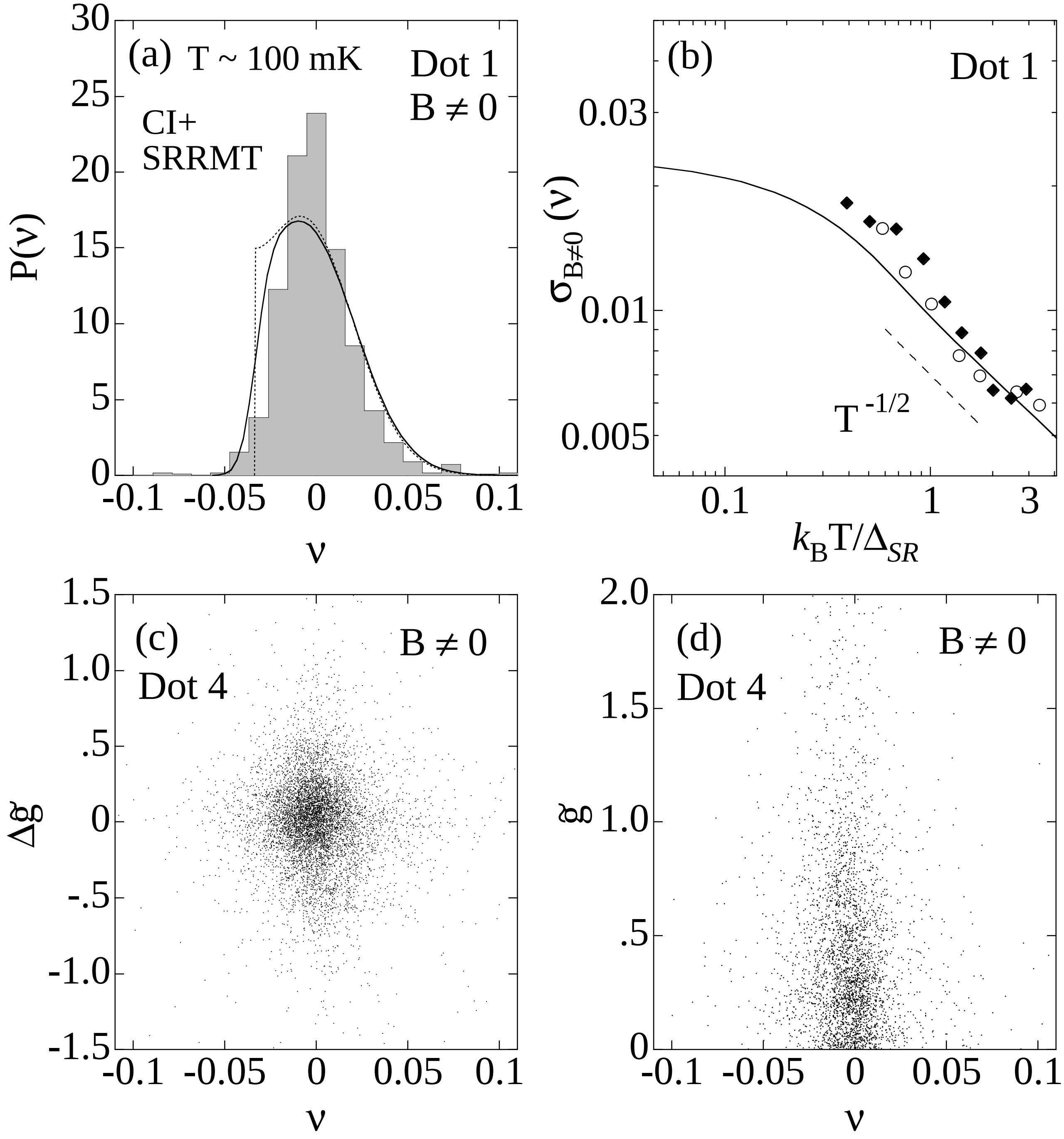

Fig. 3